\documentstyle[preprint,prd,aps]{revtex}
\begin{document}

\title{Charged particles in external fields as physical examples of \\
quasi-exactly solvable models: a unified treatment}

\author{Chun-Ming Chiang$^{1,2}$ and Choon-Lin Ho$^1$}

\address{\small \sl
$^1$Department of Physics, Tamkang University, Tamsui 25137, Taiwan\\
$^2$Kuang Wu Institute of Technology, Peitou, Taipei 112, Taiwan}

\maketitle

\begin{abstract}

We present a unified treatment of three cases of
quasi-exactly solvable problems,
namely, charged particle moving in Coulomb and magnetic
fields, for both the Schr\"odinger and the Klein-Gordon case, and
the relative motion of two charged particles in an external
oscillator potential.
We show that
all these cases are reducible to the same basic
equation,
which is quasi-exactly solvable owing to the existence of
a hidden $sl_2$ algebraic structure.
A systematic and unified algebraic solution
to the basic equation using the method of factorization is given.
Analytic expressions of the energies and
the
allowed frequencies for the three cases are given in terms of
the roots of one and the same set of Bethe ansatz equations.

\end{abstract}

\newpage

\section{Introduction}

It is well known that exact solutions are hard to come by in physics (in fact,
in all sciences).  Many exactly solvable examples presented in textbooks of
physics are only rare cases.  More often than not they serve only as paradigms
to illustrate the fundamental principles in their respective fields.  For real
problems approximation methods are indispensable.

Recently, it was found that for certain quantum-mechanical problems analytical
solutions are possible, but only for parts of the energy spectra and for
particular values of the fundamental parameters of the problems.  First it was
realized that the problem of two electrons moving in an external oscillator
potemtial was of such a class \cite{SG,Taut1}.   Later, it was
discovered that two-dimensional Schr\"odinger equations of electron moving in
an attractive/repulsive Coulomb field and a homogeneous magnetic field also
share the same characteristics [\cite{Taut2}-\cite{Taut4}].  More recently,
the latter problems were extended to the two-dimensional Klein-Gordon case
\cite{VP} and the Dirac case \cite{HoKha}.

The essential features shared by all these examples are as follows.
The differential equations (Schr\"odinger, Klein-Gordon, and
Dirac) are solved according to
the standard procedure.  One first separates out the asymptotic behaviors of
the system.  One then obtains an equation for the part which can be expanded
as a power series of the basic variable.  It is at this point that deviation
from the standard exactly solvable cases appears:  instead of the two-step
recursion relations for the coefficients of the power series so
often encountered in exactly solvable problems, one gets three-step recursion
relations.  The complexity of the recursion relations does not allow one to
do anything to guarantee normalizability of the eigenfunctions.  However, one
can impose a sufficient condition for normalizability by terminating the series
at a certain order of power of the variable; {\it i.e.} by choosing a
polynomial. By so doing one could obtain
exact solutions to the original problem, but only for certain energies and for
special values of the parameters of the problem.  For the works mentioned
above, these parameters are
the frequency of the oscillator potential and the external magnetic fields.

Soon after it was realized \cite{Tur1,HoKha} that the above quantum-mechanical
problems are just examples of the so-called quasi-exactly solvable models,
recently discovered by physicists and mathematicians [\cite{TU}-\cite{Zas}].
This is a class of quantum-mechanical problems for which several eigenstates
can be found explicitly.  The reason for such quasi-exactly solvability is
usually the existence of a hidden Lie-algebraic structure
[\cite{Tur2}-\cite{KO}].  More precisely, quasi-exactly solvable Hamiltonian
can be reduced to a quadratic combination of the generators of a Lie group
with finite-dimensional representations.

In this paper we would like to show that three of the four problems mentioned
in the second paragraph, namely, A) charged particle moving in Coulomb and
magnetic fields (Schr\"odinger
case); B) charged particle in Coulomb and magnetic fields (Klein-Gordon case);
and C) relative motion of two charged particles in an external
oscillator potential,
can be given a unified treatment.
We shall show that
all these cases are simply variations of the same basic
equation [eq.(\ref{general})
below], which is quasi-exactly solvable owing to the existence of
a hidden $sl_2$ algebraic structure.  This algebraic structure was first
realized by Turbiner for the case of two electrons in an oscillator
potential \cite{Tur1}.  We shall give a systematic and unified algebraic
solution
to the basic equation using the method of factorization presented in
\cite{HoKha} for the case A.  Our method allows
one to find the analytic expressions of the energies and the allowed
frequencies
once and for all in terms of the roots of a set of Bethe ansatz equations.
This is in sharp contrast to the method of solving recursion
relations, which must be performed for each and every order of the polynomial
part in order to get these expressions.  Our treatment also reveals that
the eigenenergies and the allowed frequencies in all these cases
are given by the roots of
the same set of Bethe ansatz equations.  This also makes explicit the close
connection between the three cases.

We will define the three problems in Sect.II.  The
basic equation underlying these three cases is then solved in Sect.III
by the method of factorization.  In Sect.IV we show how previous results,
obtained for the three cases by solving recursion relations, can be easily
reproduced from the solutions obtained in Sect.III.  The Lie-algebraic
structure underlying the basic equation is discussed in Sect.V.  Sect.VI then
concludes the paper.

\section{The three cases}

In this section we shall give a brief description of the three cases of
charged particles moving in external fields which we will consider in the rest
of the paper.  Following previous works,
we adopt the atomic units $\hbar=m=e=1$ in the CGS system.

\subsection{Electron in Coulomb and magnetic fields:
Schr\"odinger case}

This general case was considered in [\cite{Taut2}-\cite{Taut4},\cite{HoKha}].
The Hamiltonian of a planar electron in a Coulomb field and a constant magnetic
field ${\bf B}=B{\hat z}$ ($B>0$) along the $z$ direction is
\begin{equation}
H=\frac{1}{2}\left({\bf p}+\frac{1}{c}{\bf A}\right)^2 - \frac{Z}{r}~,
\end{equation}
where $c$ is the speed of light, $Z$ (positive or negative) is the charge of
the source of the Coulomb field,  and
the vector potential ${\bf A}$ is ${\bf A}=\frac{1}{2}{\bf B}\times {\bf r}$
in the symmetric gauge.

An ansatz of the eigenfunction in the polar cordinates $(r,\theta)$ is
\begin{eqnarray}
\Psi ({\bf r},t) =\frac{u(r)}{\sqrt{r}}\exp(im\theta-iEt)~,
~~~~m=0,\pm 1,\pm 2,\ldots
\label{Psi}
\end{eqnarray}
Here $m$ is the angular quantum number, and $E$ the energy.
The radial wave function $u(r)$ satisfies the radial Schr\"odinger equation
\begin{eqnarray}
\left[\frac{1}{2}\frac{d^2}{dr^2}-\frac{1}{2}\left(m^2-\frac{1}{4}\right)
\frac{1}{r^2} -\frac{1}{2}\omega_L^2 r^2 +\frac{Z}{r} + E -
m\omega_L\right]u(r)=0
\label{A}
\end{eqnarray}
where $\omega_L=B/2c$ is the Larmor frequency .

\subsection{Electron in Coulomb and magnetic fields: Klein-Gordon case}

In \cite{VP} the above problem is extended to the Klein-Gordon case, assuming
the same
ansatz of the wave function as in (\ref{Psi}).  Now the radial wave function
$u(r)$ obeys the following equation:
\begin{eqnarray}
\left[\frac{1}{2}\frac{d^2}{dr^2}-\frac{1}{2}\left(m^2-\frac{Z^2}{c^2}
-\frac{1} {4}\right)\frac{1}{r^2} -\frac{1}{2}\omega_L^2 r^2 +\frac{EZ}{c^2 r}
+ \frac{E^2}{2c^2}-\frac{c^2}{2} - m\omega_L\right]u(r)=0~.
\label{B}
\end{eqnarray}
But now, as noted in \cite{VP},  the
quantum number $m$ must satisfy the relation
\begin{equation}
m^2 - \frac{Z^2}{c^2} > 0
\end{equation}
in order for the solutions to make sense.
This relation forbids the existence of the $s$-states ($m=0$).

\subsection{Relative motion of two electrons in an external oscillator
potential}

In \cite{Taut1} the author considered the problem of
three-dimensional Schr\"odinger
equation for two electrons (interacting with Coulomb potential) in an external
harmonic-oscillator potential with frequency $\omega_{ext}$.
The Hamiltonian is
\begin{eqnarray}
H=-\frac{1}{2}\nabla^2_1+\frac{1}{2}\omega_{ext}^2 {\bf r}_1^2
-\frac{1}{2}\nabla^2_2+\frac{1}{2}\omega_{ext}^2 {\bf r}_2^2
+\frac{1}{|{\bf r}_1-{\bf r}_2|}~.
\end{eqnarray}
The total wave function is factorizable into three parts which depend
respectively only on the center of mass, the relative coordinates, and the
spins
of the electrons.  The wave function of the center of mass coordinates
satisfies the Schr\"odinger equation of a three-dimensional oscillator the
solution of which is well known.  The spin part dictates the parity of the
wave function of the relative motion.  The Schr\"odinger equation for the
relative motion is
\begin{eqnarray} \left[-\frac{1}{2}\nabla^2_r+\frac{1}{2}\omega_r^2 {\bf r}^2
+\frac{1}{2 r}\right]\phi({\bf r})=\epsilon^\prime\phi({\bf r})~,
\label{rel}
\end{eqnarray}
where ${\bf r}={\bf r}_1-{\bf r}_2$, $\omega_r =\omega_{ext}/2$, and
$\epsilon^\prime$ is one half of the eigenenergy of the relative motion (in
the notation of \cite{Taut1}).  By assuming an ansatz of the wave function in
the spherical coordinates of the form
\begin{eqnarray}
\phi({\bf r})=\frac{u(r)}{r}Y_{lm}({\hat {\bf r}})~,
\end{eqnarray}
where $Y_{lm}$ are the spherical harmonics,
we get from (\ref{rel}) the following equation
\begin{eqnarray}
\left[\frac{1}{2}\frac{d^2}{dr^2}-\frac{l(l+1)}{2}
\frac{1}{r^2} -\frac{1}{2}\omega_r^2 r^2 -\frac{1}{2r} + \epsilon^\prime
\right]u(r)=0~.
\label{C}
\end{eqnarray}

We note here that if we change the sign of the $1/r$ term in the last
equation, we get an equation that describes the
relative motion in the oscillator potential of an electron and a positron.
This case is included in our discussions.

\section{The Basic equation and the Method of factorization}

It can be seen that equations (\ref{A}), (\ref{B}) and (\ref{C}),
after an appropriate change of parameters, have the same
basic form, namely:
\begin{eqnarray}
\left[\frac{1}{2}\frac{d^2}{dr^2}-\frac{\gamma(\gamma-1)}{2}
\frac{1}{r^2} -\frac{1}{2}\omega^2 r^2  + \frac{\beta}{r} + \alpha
\right]~u(r)=0~.
\label{general}
\end{eqnarray}
Here $\beta, \gamma$ and $\omega$ ($\gamma, \omega>0$) are real
parameters, and $\alpha$ is the eigenvalue of eq.(\ref{general}).
Explicit expressions of these parameters for the three cases mentioned in the
Introduction will be given in the next section.
That this equation is quasi-exactly solvable means that, given a fixed value
of the parameter $\gamma$ and $\beta$ (or $\omega$), the equation can only be
solved exactly
only for particular set of parameter $\omega$ (or $\beta$) and eigenvalue
$\alpha$.

Now we make the following change of variables:  $x\equiv \sqrt{2\omega} r$
and $b\equiv \sqrt{2/\omega}\beta$.  Then eq.(\ref{general}) becomes:
\begin{eqnarray}
\left[\frac{d^2}{dx^2}-\frac{\gamma(\gamma-1)}{x^2}
 -\frac{x^2}{4} + \frac{b}{x} + \frac{\alpha}{\omega}
\right]~u(x)=0~.
\label{general-x}
\end{eqnarray}
The values of $\alpha$ and $b$ in
eq.(\ref{general-x}) may be found by means of a method closely
resembling the method of factorization in nonrelativistic quantum mechanics
\cite{HoKha}.  We shall discuss this method briefly below.
Let us assume
\begin{eqnarray}
u(x) =  x^\gamma\exp(-x^2/4)Q(x)~,
\label{eqff1}
\end{eqnarray}
where $Q$ is a polynomial.  As mentioned in the Introduction, the assumption
that $Q$ be a polynomial is only a sufficient, not necessary, condition
for normalizability of the eigenfunction $u(x)$.
Substituting
(\ref{eqff1}) into (\ref{general-x}), we have
\begin{eqnarray}
 \left[\frac{d^2}{dx^2} + \left(\frac{2\gamma}{x} - x\right)\frac{d}{dx} +
\left(\epsilon + \frac{b}{x}\right)\right]~Q(x) = 0~,
\label{eqrnew}
\end{eqnarray}
where $\epsilon = \alpha/\omega -(\gamma + 1/2)$.

It is seen that the problem of finding spectrum for (\ref{eqrnew})
is equivalent to determining the eigenvalues of the operator
\begin{eqnarray}
 H = - \frac{d^2}{dx^2} - \left(\frac{2\gamma}{x}-
x\right)\frac{d}{dx} - \frac{b}{x}~.
\label{hamss}
\end{eqnarray}
We want to factorize the operator (\ref{hamss})
in the form
\begin{eqnarray}
 H =  a^{+}a + p,
\label{hams1}
\end{eqnarray}
where the quantum numbers $p$ are related to the eigenvalues of
(\ref{eqrnew}) by $p=\epsilon$.
The eigenfunctions of the operator $H$ at $p=0$ must satisfy
the equation
\begin{eqnarray}
 a\psi=0~.
\label{eigenf}
\end{eqnarray}
Suppose polynomial solutions exist for (\ref{eqrnew}), say
$Q=\prod\limits_{k=1}^n
(x-x_k)$, where $x_k$ are the zeros of $Q$, and $n$ is the degree of $Q$
(we mention here that the order $n$ in this paper is equal to $(n-1)$ in
[\cite{Taut1}-\cite{VP}] where $x^{n-1}$ is the highest order term in $Q$).
Then the operator $a$ must have the form \begin{eqnarray}
 a = \frac{\partial}{\partial x} - \sum_{k=1}^n \frac{1}{x-x_k}~,
\label{oper}
\end{eqnarray}
and the operator $a^+$ has the form
\begin{eqnarray}
 a^+ = - \frac{\partial}{\partial x} - \frac{2\gamma}{x} + x -
\sum_{k=1}^n \frac{1}{x-x_k}~.
\label{conoper}
\end{eqnarray}

Substituting (\ref{oper}) and (\ref{conoper}) into (\ref{hams1}) and then
comparing the result with (\ref{hamss}), we obtain the following
set of equations for the zeros $x_k$ (the so-called Bethe ansatz
equations \cite{Ush}):
\begin{eqnarray}
 \frac{2\gamma}{x_k} - x_k - 2\sum\limits_{j\ne k}^n\frac{1}{x_j-x_k} = 0~,
\quad k=1,\ldots,n~~,
\label{Bethe}
\end{eqnarray}
as well as the two relations:
\begin{eqnarray}
b = 2\gamma\sum\limits_{k=1}^n x_k^{-1}~,\quad n=p~.
\label{relat1}
\end{eqnarray}
Summing all the $n$ equations in (\ref{Bethe}) enables us to rewrite the
first relation in (\ref{relat1}) as
\begin{eqnarray}
b = \sum\limits_{k=1}^n x_k~.
\label{relat2}
\end{eqnarray}
From $p=\epsilon$ and the second equation in (\ref{relat1}), one gets
\begin{eqnarray}
 \epsilon = n=\alpha/\omega -\left(\gamma + \frac{1}{2}\right)~.
\label{eigen}
\end{eqnarray}

For $n=1, 2$ the zeros $x_k$ and the values of the parameter $b$ for which
solutions in terms of polynomial of the corresponding degrees  exist can
easily be found from  (\ref{Bethe}) and (\ref{relat2}) in the form
\begin{eqnarray}
 n=1~,\quad x_1 &=& \pm \sqrt{2\gamma}~, \quad b = \pm \sqrt{2\gamma}~;
\phantom{mmmmmmmm} \nonumber \\
 n=2~,\quad x_1 &=& -x_2 =\sqrt{2\gamma +1}~, \quad b = 0~,
\phantom{mmmmmmmm} \nonumber \\
\phantom{mmmmm}  x_1 &=& 2\gamma/x_2~, \quad x_2 = \pm
(1+\sqrt{4\gamma+1})/\sqrt{2}~, \quad b = \pm \sqrt{2(4\gamma+1)}~.
\label{roots}
\end{eqnarray}
These expressions will be employed in the next section.

For general values of $n$ solving for the $x_k$'s from eq.(\ref{Bethe}) is
difficult, and one
must resort to numerical methods.  But some properties of the solutions are
known.  First, from eq.(\ref{Bethe}) we see that, if $\{x_k\}$
is a set of solution to (\ref{Bethe}), then so is $\{-x_k\}$.  This means, by
eq.(\ref{relat2}), that for every possible value of $b$, there is a
corresponding negative value $-b$.  Second,  as we  shall see
later in Sect.V, the number of values of $b$ for a fixed order $n$ is $n+1$.

\section{Solutions to the three cases}

We now apply the results of Sect.III to reproduce the results of previous
works.  The essential step is to solve the Bethe ansatz equations
(\ref{Bethe}) for the roots $x_k$'s for each order $n$.  Then from
eqs.(\ref{relat2}) and (\ref{eigen}) we obtain the values of the allowed pair
of frequency and energy.  It will be apparent that it is the values of $b$'s
that determine the energies and the frequencies $\omega=2\beta^2/b^2$.

\subsection{Electron in Coulomb and magnetic fields: Schr\"odinger case}

In this case, $\gamma=|m|+1/2$, $\omega=\omega_L$, $\beta=Z=\pm |Z|$, and
$\alpha=E-m\omega_L$.
The upper (lower) sign in $\beta$
corresponds to the case of attractive (repulsive) Coulomb interaction.
This will be assumed throughout the rest of this subsection.
We have
\begin{eqnarray}
\omega_L=\frac{2Z^2}{b^2}~, ~~E=\omega_L\left(n+m+|m|+1\right)~.
\label{Schro}
\end{eqnarray}
These are the general expressions for the frequency (and hence the magnetic
field) and the energy in terms of the values of $b$.

The case of $n=1$ and $n=2$ can be obtained easily.
From (\ref{roots}) and the definition of $b$ one has
\begin{eqnarray}
n=1~,\quad  \omega_L &=& 2 \frac{(Z\alpha)^2}{2|m|+1}, \quad
E_1 =  \frac{2(Z\alpha)^2}{2|m|+1}(3+m+|m|)~;\nonumber \\
n=2~, \quad \omega_L &=&  \frac{(Z\alpha)^2}{4|m|+3}, \quad
E_2 = \frac{(Z\alpha)^2}{4|m|+3}(4+m+|m|)~.
\label{frequen}
\end{eqnarray}
The corresponding polynomials are
\begin{eqnarray}
 Q_1= x-x_1 = x \mp \sqrt{2\gamma}~, \phantom{mmmmmm} \nonumber \\
 Q_2= \prod\limits_{k=1}^2 (x-x_k) = x^2 \mp x\sqrt{2(4\gamma+1)} + 2\gamma~.
\label{poly12}
\end{eqnarray}
The wave functions are described by (\ref{eqff1}).
For the repulsive Coulomb field the wave functions (for $n=1, 2$) do
not have nodes, i.e. the states described by them are ground
states, while for the attractive Coulomb field the wave function for $n=1$ has
one node (first excited state) and the wave function for $n=2$ has two nodes
(second excited state).

Let us mention here that we may consider a dual situation of the original
problem:  we may consider the magnetic field $B$ (and thus $\omega_L$) as a
fixed quantity, and eq.(\ref{Schro}) then give the allowed values of the
energy and the charge $Z$.

\subsection{Electron in Coulomb and magnetic fields: Klein-Gordon case}

For definiteness, we consider positive energy solutions for the attractive
Coulomb potential ($Z>0$).  This is the case considered in \cite{VP}. Negative
energy
solutions, and the case for repulsive Coulomb field can be treated in exactly
the same way.   In this
case, $\gamma=\sqrt{m^2-Z^2/c^2}+1/2$,
$\omega=\omega_L$, $\beta=ZE/c^2$, and $\alpha=E^2/2c^2-c^2/2-m\omega_L$.
In order for the wave function to make sense, $\gamma$ has to be real.  This
implies that $m^2-Z^2/c^2>0$, which forbids the existence of the $m=0$
states (the $s$ states) in the Klein-Gordon case, as noted in \cite{VP}.

Using $\omega_L=2\beta^2/b^2$ we get the allowed magnetic field as
\begin{eqnarray}
B=2c\omega_L=\frac{4Z^2E^2}{b^2c^3}~,
\label{omega-KG}
\end{eqnarray}
and the corresponding energy $E$ is obtained
from eq.(\ref{eigen}):
\begin{eqnarray}
n&=&\frac{\alpha}{\omega_L} -(\gamma + 1/2)\nonumber\\
&=&\frac{b^2c^2}{4Z^2}-\frac{b^2c^6}{4Z^2}\frac{1}{E^2}
-\left(1+m+\sqrt{m^2-Z^2/c^2}\right)~.
\end{eqnarray}
This equation leads to
\begin{eqnarray}
E^2=c^4\left[1-\frac{4Z^2}{b^2c^2}\left(
n+1+m+\sqrt{m^2-Z^2/c^2}\right)\right]^{-1}~.
\label{E-KG}
\end{eqnarray}
These are the most general expressions for the energy and the frequency.

To compare with the results of $n=1,2$ cases given in \cite{VP}, we substitue
these values of $n$ into (\ref{E-KG}) and (\ref{omega-KG}).  For $n=1$, we get
\begin{eqnarray}
E&=&c^2\left[1-\frac{4Z^2\left(2+m+\sqrt{m^2-Z^2/c^2}\right)}
{c^2\left(2\sqrt{m^2-Z^2/c^2}+1\right)}\right]^{-1/2}~,\nonumber\\
B&=&\frac{4E^2Z^2}{c^3\left(2\sqrt{m^2-Z^2/c^2}+1\right)}~,
\end{eqnarray}
and for $n=2$,
\begin{eqnarray}
E&=&c^2\left[1-\frac{2Z^2\left(3+m+\sqrt{m^2-Z^2/c^2}\right)}
{c^2\left(4\sqrt{m^2-Z^2/c^2}+3\right)}\right]^{-1/2}~,\nonumber\\
B&=&\frac{2E^2Z^2}{c^3\left(4\sqrt{m^2-Z^2/c^2}+3\right)}~.
\end{eqnarray}
These expressions are exactly the ones obtained in \cite{VP} by solving
recursion relations.

For negative energy solutions, the energy is given by the negative roots of
eq.(\ref{E-KG}).  The only difference is that the roots of the Bethe ansatz
equations have opposite signs, in view of eq.(\ref{relat2}).  This only changes
the nodal structure of the wave functions.  From the expression $\beta=ZE/c^2$,
we note the equivalence between the
positive (negative) energy solutions in the attractive Coulomb case and the
negative (positive) energy solutions in the repulsive Coulomb case.

Similar to case A, one may consider the dual situation in which the magnetic
field is assumed fixed, and the Bethe ansatz equations instead give the values
of the allowed pair of energy and the Coulomb charge.

\subsection{Relative motion of two electrons in an external oscillator
potential}

In this case, $\gamma=l+1/2$, $\omega=\omega_r$, $\beta=-1/2$, and
$\alpha=\epsilon^\prime$.
We have the following general solutions:
\begin{eqnarray}
\omega_r=\frac{1}{2b^2}~, ~~\epsilon^\prime
=\omega_r\left(n+l+\frac{3}{2}\right)~.
\end{eqnarray}

For the two simplest cases $n=1$ and $n=2$, we have
\begin{eqnarray}
 n=1~,\quad \omega_r=\frac{1}{4(l+1)}~, \quad \epsilon^\prime =
\frac{2l+5}{8(l+1)}~, \phantom{mmmmmmmm} \nonumber \\
 n=2~,\quad \omega_r=\frac{1}{4(4l+5)}~, \quad \epsilon^\prime =
\frac{2l+7}{8(4l+5)}~, \phantom{mmmmmmmm} \nonumber \\
\end{eqnarray}
These are exactly the expressions given in \cite{Taut1}.
They are also the solutions (for $n=1,2$) for the case of an electron and a
positron in the oscillator potential ($\beta=+1/2$).  The nodal structures of
the wave functions are the same as those described for case A.

\section{Hidden Lie-algebraic structure of the basic equation}

The basic equation (\ref{general}), or its equivalent form
(\ref{eqrnew}), possesses an underlying Lie-algebraic structure that is
responsible for its quasi-exactly solvability.

In fact, Turbiner has indentified a $sl_2$ structure for the case of two
charged particles in an oscillator potential \cite{Tur1}. In view of the fact
that
all the previous cases considered in this papers are related to the same
basic equation (\ref{general}), one expects
the same hidden structure to be present in all these cases.  This is indeed
the case, and it is sufficient to show that $sl_2$ algebra is in fact the
underlying structure possessed by (\ref{general}) or (\ref{eqrnew}).
In this section we shall carry out Turbiner's analysis to
eq.(\ref{eqrnew}),
with only slight modifications in the parameters to suit the general situation.
Only the main ideas are given here, and we refer the reader to \cite{Tur1} for
details.

Let us construct three generators in the following manner:
\begin{eqnarray}
J_n^+ &=& r^2 \frac{d}{dr} - n r~,\nonumber\\
J_n^0 &=& r \frac{d}{dr} - \frac{n}{2}~,\nonumber\\
J_n^- &=& \frac{d}{dr}~.
\label{gen}
\end{eqnarray}
These generators realize the $sl_2$ algebra:
\begin{eqnarray}
\left[J_n^+,J_n^-\right]=-2J_n^0~,~~~\left[J_n^0,J_n^\pm\right]=\pm J_n^\pm
\end{eqnarray}
for any value of the parameter $n$.  If $n$ is a non-negative integer, then
there exists for the $sl_2$ algebra a $(n+1)$-dimensional irreducible
representation ${\cal P}_{n+1}(r)=\langle 1,r,r^2,\ldots,r^n\rangle$.
From this it is clear that any differential operator formed by taking
polynomial of the generators (\ref{gen}) will have the space ${\cal P}_{n+1}$
as the finite-dimensional invariant subspace.  This is the main idea
underlying the quasi-exactly-solvable operators [\cite{Tur1}-\cite{KO}].

Now consider the quasi-exactly-solvable operator which is quadratic in the
$J_n$'s:
\begin{eqnarray}
T_2=-J_n^0J_n^- + 2\omega J_n^+-\left(\frac{n}{2}+2\gamma\right)J_n^-
- \omega n ~.
\label{T2-1}
\end{eqnarray}
This operator belongs to the class VIII according to the classification given
in \cite{Tur2}.  In terms of $r$, $T_2$ becomes
\begin{eqnarray}
T_2=-r\frac{d^2}{dr^2} + 2\left(\omega
r^2-\gamma\right)\frac{d}{dr} -2\omega nr ~.
\label{T2-2}
\end{eqnarray}

Let us now consider the eigenvalue problem
\begin{eqnarray}
T_2 Q(r) =2\beta (n) Q(r)~.
\label{T2-eigen}
\end{eqnarray}
This eigenvalue problem possesses $n+1$ eigenvalues $\beta(n)$,
and the corrsponding eigenfunctions are in the form of
polynomial of the $n$-th power, while other eigenfunctions are non-polynomial
which in general cannot be found in closed analytic form \cite{Tur1}.
Let us first substitute the form (\ref{T2-2}) into eq.(\ref{T2-eigen}), then
divide the resulting equation by $r/2$, we arrive at
\begin{eqnarray}
\left[\frac{1}{2}\frac{d^2}{dr^2}-\left(\omega
r-\frac{\gamma}{r}\right)\frac{d}{dr} +\omega n
+\frac{\beta(n)}{r}\right]~Q(r)=0~.
\label{e1}
\end{eqnarray}
Finally, we divide (\ref{e1}) by $\omega$ and change the variable $r$ to
$x\equiv \sqrt{2\omega}r$.  This leads to the following equation
\begin{eqnarray}
\left[\frac{d^2}{dx^2} + \left(\frac{2\gamma}{x} - x\right)\frac{d}{dx} +
\left(n + \sqrt{\frac{2}{\omega}}\beta(n)\frac{1}{x}\right)\right]Q(x)
= 0~.
\label{e2}
\end{eqnarray}
This is exactly eq.(\ref{eqrnew}), provided that $\epsilon=n$ and
$b=\sqrt{2/\omega}\beta(n)$.
This means that
eq.(\ref{eqrnew})
is quasi-exactly solvable if $\epsilon=n$, which is exactly our relation in
eq.(\ref{eigen}), and that there are only $n+1$ allowed values of
$b=\sqrt{2/\omega}\beta$ in
eq.(\ref{eqrnew}) [{\it cf.} eq.~(\ref{roots})].

Translating back to the original
three cases considered in this papepr, these results imply the following.  In
case C,  $\beta$ is a fixed parameter ($\beta=\pm 1/2$), hence
the finite number ($=n+1$) of the values of $b$ implies the same number of
the allowed frequency $\omega_{ext}$ of the external oscillator potential and
the corresponding energy.  This
is the case found in \cite{Taut1} and presented here again from a new light.
For case A ($\beta=Z$) and case B ($\beta=ZE/c^2$), the above results mean
that, at a fixed order $n$, there are exactly $n+1$
allowed values of the pair of energy and magnetic field
for a fixed Coulomb charge, or of the pair of energy and
Coulomb charge for a fixed magnetic field.

Furthermore, it has been shown \cite{Tur1} that
there exist $[(n+1)/2]$ positive eigenvalues and the same number of negative
eigenvalues of the $b$ (here $[a]$ represents the integral part of $a$).
In the general situation considered in this paper, positive (negative) values
of $b$ correspond to the attractive (repulsive) Coulomb
field for positive energy solutions.  For negative energy solutions, the sign
of $b$ is reversed for the two kinds of Coulomb field.  Hence, our
unified treatment together with the Lie-algebraic analysis of these
cases give a very simple explanation as to why the number of the
positive
energy levels for a fixed order $n$ considered in [\cite{Taut1}-\cite{VP}] are
all equal to $[(n+1)/2]$.

\section{Conclusions}

In this paper we have presented a unified treatment of three cases of
quasi-exactly solvable problems,
namely, charged particle moving in Coulomb and magnetic
fields, for both the Schr\"odinger and the Klein-Gordon case, and
the relative motion of two charged particles in an external
oscillator potential.
We show that
all these cases are reducible to the same basic
equation [eq.(\ref{general})],
which is quasi-exactly solvable owing to the existence of
a hidden $sl_2$ algebraic structure.
A systematic and unified algebraic solution
to the basic equation using the method of factorization is given.
Our method allows one to express the analytic expressions of the energies and
the
allowed frequencies once and for all in terms of the roots of a set of Bethe
ansatz equations.   Our treatment also reveals that
the eigenenergies and the allowed frequencies in these cases
are all given by the roots of
the same set of Bethe ansatz equations.

\vskip 3cm
\centerline{\bf Acknowledgment}

This work was supported in part by the Republic of China through Grant No.
NSC 89-2112-M-032-020.


\end{document}